\documentclass[apl,groupedaddress,twocolumn,preprintnumbers,amsmath,amssymb]{revtex4}
\usepackage{graphicx}
\usepackage{dcolumn}
\usepackage{bm}

\begin{document}

\title{Reduction of spin transfer by synthetic antiferromagnets}

\author{N. C. Emley}
\author{F. J. Albert}
\author{E. M. Ryan}
\author{I. N. Krivorotov}
\author{D. C. Ralph}
\author{R. A. Buhrman}

\affiliation{Cornell University, Ithaca, NY 14853-2501}

\author{J. M. Daughton}
\author{A. Jander}

\affiliation{NVE Corporation, Eden Prairie, MN 55344}

\pacs{73.40.-c, 75.70.Cn, 75.60.Jk}

\date{\today}

\begin{abstract}
Synthetic antiferromagnetic layers (SAF) are incorporated into
spin transfer nanopillars giving a layer composition
[Co$_{bottom}$/Ru/Co$_{fixed}$]/Cu/Co$_{free}$, where square
brackets indicate the SAF.  The Co$_{bottom}$ and Co$_{fixed}$
layers are aligned antiparallel (AP) by strong indirect exchange
coupling through the Ru spacer.  All three magnetic layers are
patterned, so this AP alignment reduces undesirable dipole fields
on the Co$_{free}$ layer. Adding the Co$_{bottom}$/Ru layers
reduces the spin polarization of the electron current passing
through the nanopillar, leading to a decreased spin-torque per
unit current incident on the Co$_{free}$ layer. This may be
advantageous for device applications requiring a reduction of the
effects of a spin-torque, such as nanoscale CPP-GMR read heads.
\end{abstract}

\maketitle

The reversal of a thin ferromagnetic layer by application of a
spin-polarized current, or spin transfer effect (ST), has been
extensively studied in systems with the familiar
Co$_{fixed}$/Cu/Co$_{free}$ current-perpendicular-to-plane (CPP)
pseudo-spin valve composition
\cite{katineprl,albertapl,grollier,wegrowe} as well as other
magnetic trilayers \cite{sun_apl1,urazhdin}. The prevailing
theories \cite{berger,slonczewski} indicate that the
spin-polarized current applies a spin-torque, local to the
Cu/Co$_{free}$ interface, that can induce a dynamical response
from the Co$_{free}$ magnetization. Such dynamics, although
important for the study of ST, are parasitic for more passive
devices such as CPP giant magnetoresistance (GMR) read heads,
where the Co$_{free}$ layer is sensitive to stray fields from
magnetic bits on a hard drive medium \cite{nagasaka}. ST-induced
dynamics would give erroneous signals in nanoscale devices and so
it is advantageous to minimize the effects of a spin-torque in
such devices.

In this letter we present the results of an experiment where a
third, oppositely aligned magnetic layer (Co$_{bottom}$) has been
incorporated into the CPP spin valve structure adjacent to the
Co$_{fixed}$ layer. We investigate field \emph{H} and a DC current
\emph{I}-induced switching of the Co$_{free}$ layer in structures
with layer composition
Cu(100)/Co$_{bottom}$(11.5)/Ru(0.7)/Co$_{fixed}$(8)/Cu(6)
/Co$_{free}$(2)/Cu(2)/Pt(30) (in nm), where all three Co layers
are patterned in a nanopillar geometry. Interlayer exchange
coupling through the thin Ru spacer gives a strong AP alignment of
the two adjacent Co layers \cite{parkin_mauri}. The
Co$_{bottom}$/Ru/Co$_{fixed}$ trilayer thus forms a synthetic
antiferromagnet (SAF), where magnetostatic fields from the two Co
layers are in opposition and the overall dipolar coupling to the
Co$_{free}$ layer is reduced.

\begin{figure}
\includegraphics[width=8.5cm]{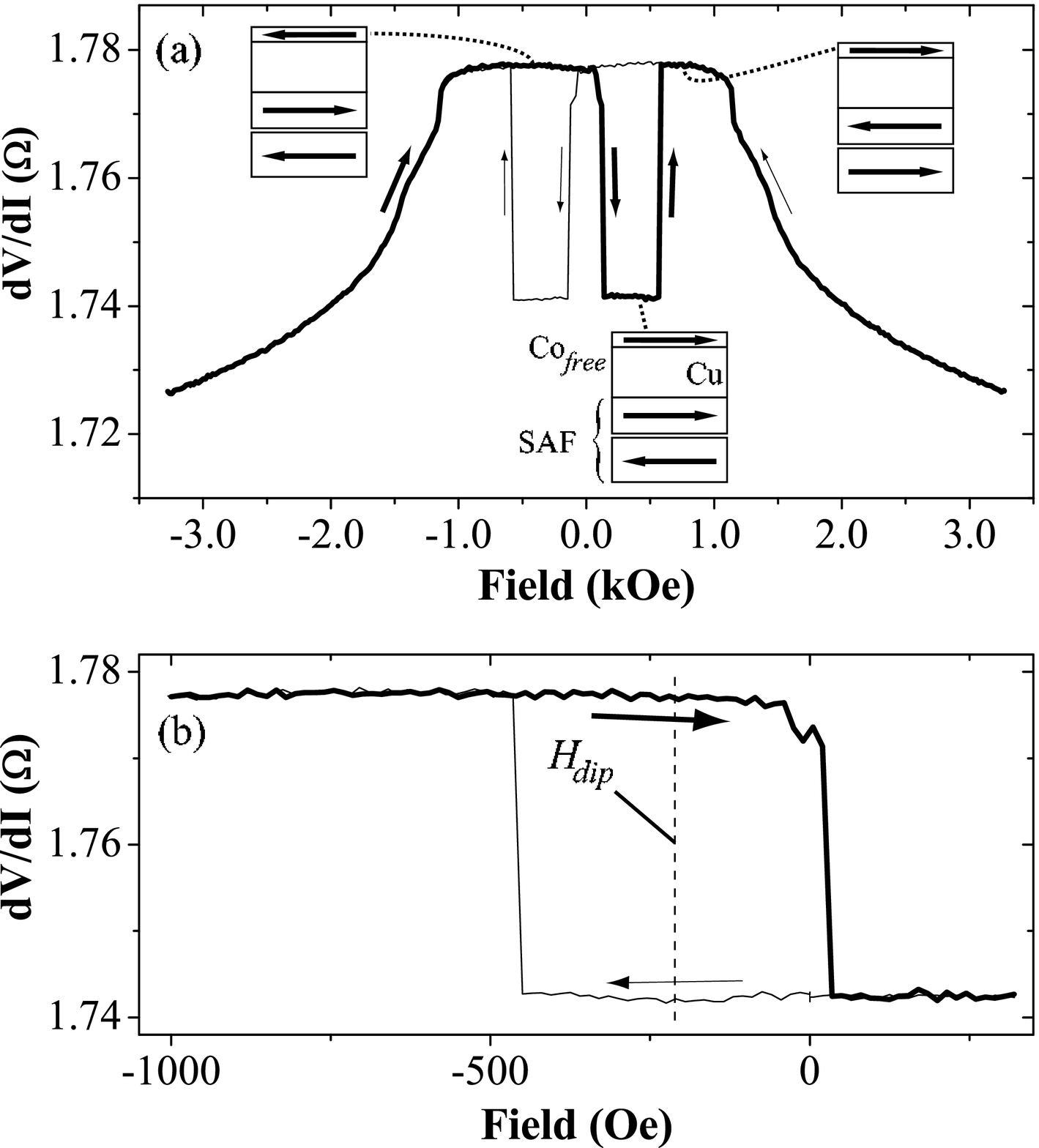}
\caption{(a) High-field GMR of the SAF nanopillar. Simulated
devices are shown to indicate the alignment of the three
magnetizations as the field is swept from negative to positive
values. (b) Low-field GMR of the same device. The field is scanned
asymmetrically to isolate the switching of the Co$_{free}$ layer.
The dipolar field ($H_{dip}^{\textrm{SAF}} \approx -220$ Oe) is
taken as the offset of the hysteresis loop, indicated by the
dashed line.}
\end{figure}

All layers are DC sputter-deposited in a 1 mTorr Ar ambient onto
thermally oxidized Si wafer substrates. Base pressures are $\leq 3
\times 10^{-8}$ Torr. Electron beam lithography, reactive ion
etching, and evaporation are used to define a mask which protects
the underlying layers during an ion beam etch step. The etching is
timed to stop partway through the thick Cu buffer, patterning all
three Co layers. SiO$_{2}$ is deposited with a plasma enhanced
chemical vapor deposition process. Photolithography, subsequent
ion beam etching steps, and sputter deposition define top and
bottom leads in a 4-point CPP configuration. Resistance
measurements are made at \emph{T} = 295 K using a Wheatstone
bridge and lock-in amplifier technique with a 25 $\mu$A excitation
current $i_{ex}$.  Here, negative \emph{I} indicates electron flow
from the SAF to the Co$_{free}$ layer.

Fig. 1(a) shows the device GMR (\emph{I} = 0) with \emph{H}
applied in-plane.  The continuing decrease in \emph{dV/dI} at the
maximum \emph{H} is the gradual breaking of the SAF AP alignment
\cite{parkin_more}.  To distinguish between switching events for
$|H| < 1.0$ kOe, we use a Stoner-Wolfarth simulation where total
energy (Zeeman, anisotropy, interlayer exchange, and dipole field)
is minimized for all three layers at each 4 Oe increment in
\emph{H}. This simulation confirms the different magnetic
configurations, shown pictorially in Fig 1(a). For $H \approx 150$
Oe the switch from high to an intermediate resistance state
corresponds to the reversal of the Co$_{free}$ layer, while the
jump back to high resistance at $H \approx 600$ Oe is the entire
SAF layer switching in unison.

In Fig. 1(b) the magnetic field is scanned over an asymmetric
range, -1000 Oe $<$ \emph{H} $< 300$ Oe, in order to isolate the
switching of the Co$_{free}$ layer. The offset of this hysteresis
loop is taken as the dipole field on the Co$_{free}$ layer
$|H_{dip}^{\textrm{SAF}}| \approx 220$ Oe. The two thicknesses of
the SAF magnetic layers are chosen specifically to minimize the
combined dipole field halfway through the Co$_{free}$ layer.
Imperfections in the magnetic layers during fabrication are most
likely responsible for $H_{dip}^{\textrm{SAF}} \neq 0$. Dipole
field calculations from surface currents on an isolated magnetic
disk (i.e. no SAF pair) show $|H_{dip}| \approx 400$ Oe halfway
through the Co$_{free}$ layer. The resistance changes in Fig. 1(b)
shift in \emph{H} as a bias \emph{I} is applied (shown in Fig. 3),
a further indication that the minor loop is the Co$_{free}$ layer
switching since the SAF is too thick to be affected by the
spin-torque \cite{albertprl}.

Looking at the resistance area product ($\Delta R\cdot A$) from
the GMR of 35 SAF samples, we find $(\Delta R\cdot
A)_{\textrm{SAF}}$ = $0.45 \pm 0.07$ $m\Omega \cdot \mu m^{2}$.
For 59 normal samples without the Co$_{bottom}$/Ru layers but with
identical thicknesses for the rest of the trilayer, we measure
$(\Delta R\cdot A)_{\textrm{normal}} = 0.94 \pm 0.19$ $m\Omega
\cdot \mu m^{2}$, almost a factor of 2 larger. This reduction of
$\Delta R\cdot A$ for the SAF samples is attributed to the reduced
polarization of the electrons that pass through and are reflected
from the SAF trilayer compared to the case of a single Co fixed
layer. Both the Co$_{bottom}$ and the Co$_{fixed}$ layers in the
SAF are considerably thinner than the room temperature
spin-diffusion length of Co ($\ell^{\textrm{Co}}_{sf} \approx 38$
nm \cite{piraux}), and the Ru coupling layer is also much thinner
(0.7nm) than its spin-diffusion length ($\sim$14 nm \cite{eid}).
Consequently, all of the interfaces of the SAF play a role in the
spin-filtering \cite{upadhyay} and collectively determine the net
spin polarization of the current that impinges onto the
Co$_{free}$ layer in these near-ballistic ST devices.

While the spin-filtering that results from the electronic
structure of the two Co/Ru interfaces \cite{rampe,zhong} and any
bulk spin-dependent scattering that does occur can be expected to
modify the effect, the two oppositely aligned magnetizations of
the SAF pair will clearly reduce the spin current amplitude that
passes through or, depending on the bias current direction,
reflects off the SAF.  Since the magnetoresistance signal $\Delta
R\cdot A$ is, in the ballistic limit, linearly dependent upon the
effective polarization $\eta_{\textrm{eff}}$ of this current, the
reduced magnetoresistance signal from SAF devices indicates that
$\eta_{\textrm{eff}} \approx \frac{1}{2}\eta_{\textrm{Co}}$ where
$\eta_{\textrm{Co}}$ is the polarization produced by the spin
filtering of a single fixed Co layer.

We note that the $\Delta R\cdot A$ for normal samples here is
larger than for those reported in \cite{albertprl}. We suspect
that this difference is due to the fact that the samples here and
those in \cite{albertprl} were prepared in different sputter
systems which yield multilayer films with different interfacial
qualities. The Co layers in this study had 37\% larger grain sizes
(from x ray diffraction measurements) and larger rms interfacial
roughness ($\sim3 \times$, from AFM measurements) than those in
\cite{albertprl}. A detailed understanding of the role of
interfacial quality on $\Delta R\cdot A$ is still lacking,
however.

\begin{figure}
\includegraphics[width=8.5cm]{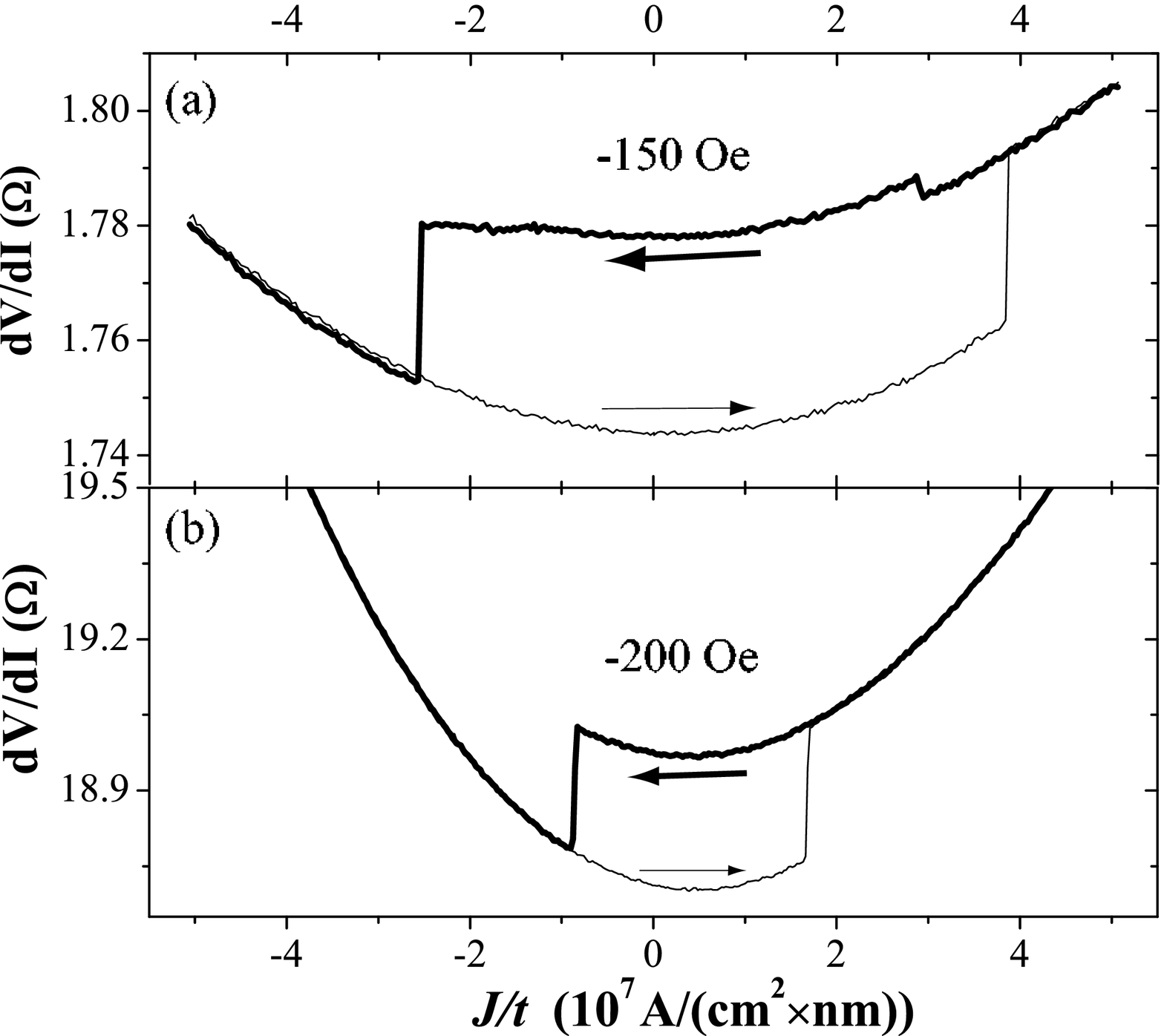}
\caption{dV/dI versus current density normalized to the free layer
thickness \emph{J/t} for (a) a 90 $\times$ 190 nm elliptical SAF
sample and (b) a 70 $\times$ 130 nm elliptical normal sample.  The
resistance values in (b), a two-point measurement, include lead
resistance $\sim$6 $\Omega$ and contact resistance $\sim$9
$\Omega$.}
\end{figure}

Not surprisingly, we find that the ST switching is also
susceptible to the reduced $\eta_{\textrm{eff}}$ from the SAF. In
Fig. 2(a) we show a ST scan for a SAF sample at low field
(\emph{I} ramp rate = 0.5 mA/sec) and a similar scan from a normal
sample (1.0 mA/sec)
Co$_{fixed}$(40)/Cu(10)/Co$_{free}$(3)/Cu(2)/Pt(30) (in nm), where
the Co$_{fixed}$ layer is unpatterned, in Fig. 2(b).  We plot the
current density \emph{J} normalized to the Co$_{free}$ layer
thickness \emph{t} because this is the value most directly related
to the spin-torque \cite{albertprl}.

For four SAF samples, we measure $\Delta J_{c}/t = 7 \pm 1 \times
10^{7}$ A/(cm$^{2} \times$ nm), while for 24 normal samples,
$\Delta J_{c}/t = 3.0 \pm 1.0 \times 10^{7}$ A/(cm$^{2} \times$
nm), an increase by a factor of $\sim$2.3. Here, $\Delta J_{c}
\equiv J^{+}_{c} - J^{-}_{c}$, where $J^{+}_{c}$ ($J^{-}_{c}$) is
the critical current density for switching Co$_{free}$ P to AP (AP
to P) with Co$_{fixed}$. There is a small difference in the Cu
spacer thickness between the SAF and normal devices, but this
would only account for a 2\% change in $\eta_{\textrm{eff}}$,
which is well within our uncertainties.  The polarization of the
conduction electrons that exert a spin-torque on the Co$_{free}$
layer may depend on the direction of the current flow. For
$J^{-}$, electrons traverse the fixed layer (single Co or SAF) and
are thereby spin filtered to produce a current with polarization
$\eta_{-}$ that impinges on the Co$_{free}$ layer. For $J^{+}$,
the incident electrons that are spin-filtered by the Co$_{free}$
layer traverse the Cu spacer and impinge onto the fixed layer.
From there a portion are reflected back to the Co$_{free}$ layer,
after being re-polarized $\eta_{+}$ by the spin-filtering effects
of the fixed layer (single Co or SAF), and exert a spin-torque on
the Co$_{free}$ layer. For simplicity we assume that the effective
polarization of the electron current exerting a spin-torque on the
Co$_{free}$ layer is the same in both cases, $\eta_{\textrm{eff}}
= \eta_{+} = \eta_{-}$.

From \cite{slonczewski}, $\Delta J_{c}/t \propto \alpha
[g(0,\eta)^{-1} + g(\pi,\eta)^{-1}]$. Here $\alpha$ is the Gilbert
damping parameter and $g(\theta,\eta)$ is a measure of the
spin-torque that is exerted on the free layer as a function of its
alignment with the fixed layer and is a monotonically increasing
function of $\eta$. Assuming $\eta_{\textrm{eff}} \approx 0.4$ and
$0.2$ for normal and SAF devices, respectively, and that $\alpha$
is the same for both types of devices, we plug these
$\eta_{\textrm{eff}}$ values into the $g(\theta,\eta)$ expression
derived by Slonczewski \cite{slonczewski} and find $\frac{(\Delta
J_{c}/t)_\textrm{SAF}}{(\Delta J_{c}/t)_\textrm{normal}} \approx$
2.5, in reasonable agreement with the data.

\begin{figure}
\includegraphics[width=8.5cm]{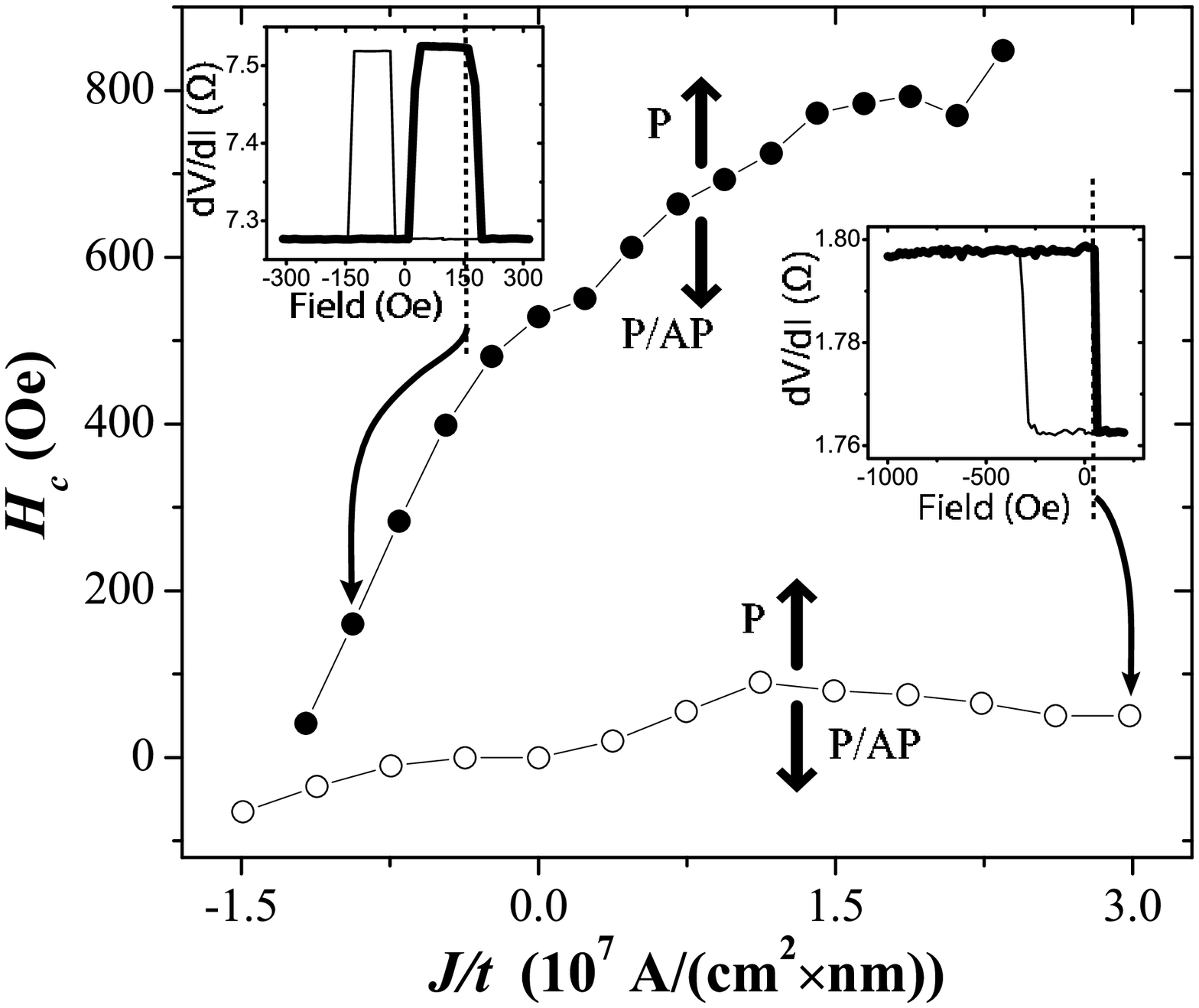}
\caption{$H_{c}$ plotted versus $J/t$, the current density
normalized to the Co$_{free}$ layer thickness, for normal
($\bullet$) and SAF ($\circ$) samples.  The data illustrate the
boundary separating the bistable P/AP region and the stable P
region as shown. Insets show how $H_{c}$ was measured.}
\end{figure}

We investigate the dependence of the ST \emph{I}-\emph{H} phase
diagram on $\eta_{\textrm{eff}}$ by measuring the coercivity
$H_{c}$ of the Co$_{free}$ layer as a function of $J/t$ for both
normal and SAF samples, shown in Fig. 3. The normal samples (same
as those shown in Fig. 2(b)) have an unpatterned Co$_{bottom}$
layer which has a naturally smaller $H_{c}$, making it simpler to
identify the Co$_{free}$ layer switching. These plots mark the
respective boundaries between the bistable P/AP and P regions, as
measured in other experiments with non-SAF samples
\cite{myers2,grollier,kiselev}. The important point of Fig 3. is
that the slope of $H_{c}$ vs. $J/t$ is much larger for normal
samples than for SAF samples, highlighting the weaker influence,
on the Co$_{free}$ reversal, of the reduced $\eta_{\textrm{eff}}$
in SAF samples. Spin-torque-induced excitations of a nanomagnet
have been described by thermal activation models where either the
effective barrier height or the temperature is modified by
$\eta_{\textrm{eff}}$ \cite{li_zhang,koch_sun,urazhdin} and so a
reduced $\eta_{\textrm{eff}}$ correspondingly has a weaker
influence on the activation process.

In summary, we have added Co$_{bottom}$/Ru layers to the familiar
Co$_{fixed}$/Cu/Co$_{free}$ CPP magnetic nanopillar composition.
The Co$_{bottom}$ and Co$_{fixed}$ layers are AP due to exchange
coupling through the Ru spacer and succeed in reducing unfavorable
dipole fields on the Co$_{free}$ layer.  It is clear that these AP
magnetic layers also reduce the spin polarization of \emph{I} from
the bulk Co value or single Co spin-filter value. Such reduction
of the current polarization is advantageous for nanoscale devices
seeking to reduce the effects of a spin-torque, such as CPP-GMR
read heads \cite{nagasaka}, where the reduction in $\Delta R\cdot
A$ due to the SAF can be countered by partially oxidizing the
magnetic interfaces \cite{nagasaka_jap}.

This work was supported by the NSF through the NSEC program.
Fabrication was done at the Cornell Nanofabrication Facility which
is a node of the NSF-supported National Nanofabrication Users
Network.

\bibliography{Bottom_SAF_APLbib}
\end{document}